\documentclass[superscriptaddress,showpacs,aps,twocolumn,prb,floatfix]{revtex4-1}
\usepackage{amssymb}
\usepackage{amsmath}
\usepackage[dvips]{graphicx}
\usepackage{bm}

\begin{document}
\title{Effects of spin density wave quantization on the electrical transport in epitaxial Cr thin films}
\author{E. Osquiguil}
\author{E. E. Kaul}
\author{L. Tosi}
\author{C. A. Balseiro}
\affiliation{Centro At\'omico Bariloche and Instituto Balseiro, Comisi\'on Nacional de
Energ\'{\i}a At\'omica, 8400 Bariloche, Argentina}

\date{\today }

\begin{abstract}

We present measurements of the electrical resistivity, $\rho$, in epitaxial Cr films of different thicknesses grown on MgO (100) substrates, as a function of temperature, $T$. The $\rho(T)$ curves display hysteretic behavior in certain temperature range, which is film thickness dependent.  The hysteresis are related to the confinement of quantized incommensurate spin density waves (ISDW) in the film thickness.  Our important finding is to experimentally show that the temperature $T_{mid}$ where the ISDW changes from $N$ to $N$\,+\,1 nodes {\it decreases} as the film thickness {\it increases}.
Identifying $T_{mid}$ with a first order transition between ISDW states with $N$ and $N$\,+\,1 nodes,  and using a Landau approach to the free energy of the ISDW together with Monte Carlo simulations, we show that the system at high temperatures explores all available modes for the ISDW, freezing out in one particular mode at a transition temperature that indeed decreases with film thickness, $L$. The detailed dependence of $T_{mid}(L)$ seems to depend rather strongly on the boundary conditions at the Cr film interfaces.

\end{abstract}

\maketitle
\section{Introduction}

Besides the superconducting state, itinerant antiferromagnetism is 
another macroscopic manifestation of a collective quantum state of the 
charged electrons in a metal\cite{overhauser}. Bulk Cr is the 
paradigmatic and unique simple metal showing this electronic ground state. In single crystalline samples, an incommensurate spin density wave (ISDW) forms below
the N\'eel temperature $T_N$\,=\,311\,K,
characterized by wave vectors ${\mathbf Q_\pm}$\,=\,$(2\pi/a_{Cr})(1\pm\delta)$ along the $\langle 100\rangle$ directions\cite{fawcett}. Here $a_{Cr}$\,=\,2.88\,\AA\ is the lattice parameter of the Cr bcc unit cell
and $\delta$\,=\,$a_{Cr}/\Lambda_{ISDW}$ is the deviation from commensurability, with $\Lambda_{ISDW}$ the wavelength of the ISDW. This deviation is temperature dependent,\cite{werner} leading
to a variation of $\Lambda_{ISDW}$ from 60\,\AA\ at 10\,K to 78\,\AA\- at 310\,K.
Above the spin-flip temperature $T_{sf}$\,=\,123\,K, the spin waves are transverse with the spin ${\mathbf S}$ perpendicular to ${\mathbf Q}$.
Contrary to what happens in single crystals in which all three crystallographic directions are equivalent, in thin Cr films\cite{zabel} the wave vector ${\mathbf Q}$ orients perpendicular to the film surface as its thickness decreases\cite{sonntag,bodeker}, and the ISDW's are transverse for all temperatures\cite{bader} below $T_N$. This confinement of the ISDW leads to the quantization of the wave vector ${\mathbf Q}_N$, with $N$ an integer, giving rise to very interesting hysteretic phenomena.

Fullerton et al.\cite{fullerton} measured neutron scattering and electrical transport in Cr/Cr$_{97.5}$Mn$_{2.5}$ (001) superlattices, where the CrMn layers, being antiferromagnetic with a higher $T_{N}$, exert some stringent boundary conditions on the Cr surfaces. They found that temperature cycling produced irreversibility in the derivative of the resistivity and in the neutron scattering scans. These features were attributed to a change in an odd number of nodes in the ISDW inside the Cr layers which is compatible with the magnetic coupling at the interfaces with the CrMn layers.

Recently, Kummamuru et al.\cite{kummamuru} reported on transport 
results in epitaxially grown Cr (100) thin films of different 
thicknesses. Here the boundary conditions are different from those in 
the superlattices of Fullerton et al. However, they found thermal 
hysteretic behavior of the Hall coefficient and resistivity for 
thicknesses below 510\,\AA. They interpreted these results in terms of 
ISDW domains with $N$ nodes (at high temperatures) or $N$+1 nodes (at 
low temperatures) in the Cr film. For thicker films (3500\,\AA) this 
mechanism disappears, and only an irreversibility in the $\rho (T)$ 
curves was seen in a wider temperature range, and explained in terms of domain-wall scattering of electrons.

If hysteretic phenomena are indeed related to a first order phase transition taking place between domains of ISDW's with $N$ and $N$+1 nodes inside the Cr layers, it is natural to ask how the transition temperature depends on the Cr thickness. Since for thicker films the number of available modes for the ISDW increases in the whole temperature range where $\Lambda_{ISDW} (T)$ changes, one would naively expect to find the hysteretic behavior at higher temperatures in the thicker films. In this article we address these issues both experimentally and theoretically, and demonstrate that exactly the opposite behavior is found, i.e. the hysteretic region and the transition temperature move towards lower temperatures as the film thickness is increased. We therefore believe that this work throws new light on the understanding of the behavior of ISDWs in confined geometries.

\section{Experiment}

Cr thin films were grown epitaxially on MgO (100) substrates using DC 
Magnetron Sputtering in a similar way as that reported by Kummamuru et 
al..\cite{kummamuru} The films, with thicknesses between  270\AA \- and 
1100\AA, were characterized by x-ray diffraction showing rocking curves 
that have an angular dispersion at FWHM around the [002] peak of $0.5^\circ$, indicating a very good degree of epitaxy. Results from AFM scans show a mean surface roughness of about 20\AA, which is the third part of $\Lambda_{ISDW}$ at low temperatures. All films were patterned using photolithography and chemical etching in a convenient configuration to perform four terminal measurements of the resistance. It was measured as a function of $T$ between 15 and 340\-K in a commercial cryo-cooler. The error in the resistance measurement was 1/10000, almost independent of $T$.

\section{Results and Discussion}

\begin{figure}[tbp]
\includegraphics[width=0.45\textwidth]{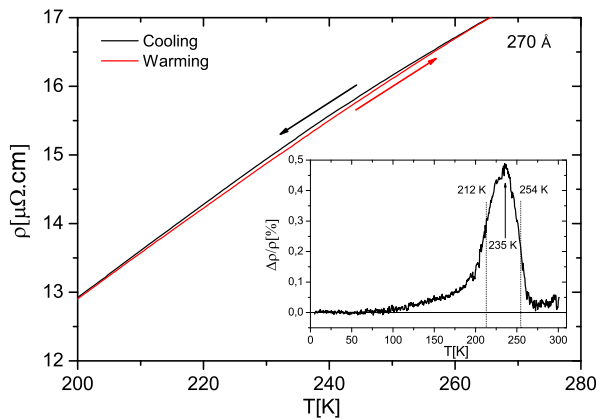}
\caption{(Color online) Resistivity as a function of temperature for a cooling-warming cycle for the 270\,\AA \- film. The inset shows the difference $(\rho_{c} - \rho_{w})/\rho_{w}$, expressed as a percentage. }
\label{rho}
\end{figure}

In Fig.\ref{rho} we show the resistivity of the 270\,\AA \- film as a function of temperature in the temperature region where the hysteretic behavior appears in $\rho(T)$ while making a cooling-warming cycle.  The inset shows the difference
between $\rho_{cooling}$ and $\rho_{warming}$ ($\rho_{c}-\rho_{w}$) normalized by $\rho_{w}$
and expressed as a percentage. Clearly, in a rather small temperature window, the resistivity while cooling down is higher than the values of $\rho (T)$ when warming up the film by an amount that reaches 0.5\% at it's maximum. This value is of the same order of magnitude than that reported by Kummamuru et al.\cite{kummamuru}

\begin{figure}[tbp]
\includegraphics[width=0.45\textwidth]{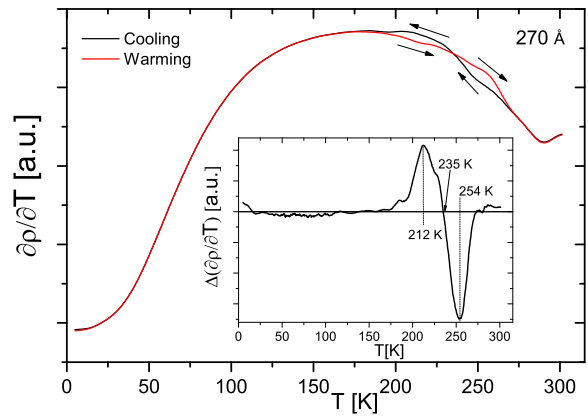}
\caption{(Color online) Derivatives of the $\rho (T)$ curves shown in 
Fig. \ref{rho}. The inset shows the difference of $\partial\rho/\partial T$ between the cooling and warming cycles.}
\label{der}
\end{figure}

We observed that the derivative of the resistivity curves has an intrinsic noise in the temperature region where the hysteresis is seen, which is not related to the measurement method. We will address this issue in a forthcoming article.\cite{ruido}
For the purpose of the physics we want to analyze here, it is enough to 
work with the smoothed curves of $\partial\rho/\partial T$ as a 
function of T as shown in Fig.\ref{der}, corresponding to the data 
displayed in Fig.\ref{rho}. Clearly, the hysteretic behavior seen in $\rho (T)$
is reflected also in the derivatives of both, the cooling and warming cycles.
The inset shows the difference between them. We define the temperature 
at which the derivatives are equal within the hysteretic zone (marked 
with an arrow) as $T_{mid}$(=\,235\,K for 270\,\AA). This coincides 
with the temperature at the maximum of $(\rho_{c} - 
\rho_{w})/\rho_{w}$. The temperatures where the difference between the 
derivatives has its maximum and minimum values, below and above 
$T_{mid}$, are marked in the insets of Figs.\ref{rho} and \ref{der} 
using dotted vertical lines. We call these temperatures $T_L$ and 
$T_H$ (212\,K and 254\,K respectively, for 270\,\AA) and define $\Delta T = T_H - T_L$. As we will show below, the $\Delta T$ region where the hysteresis occurs is film thickness dependent.

\begin{figure}[tbp]
\includegraphics[width=0.45\textwidth]{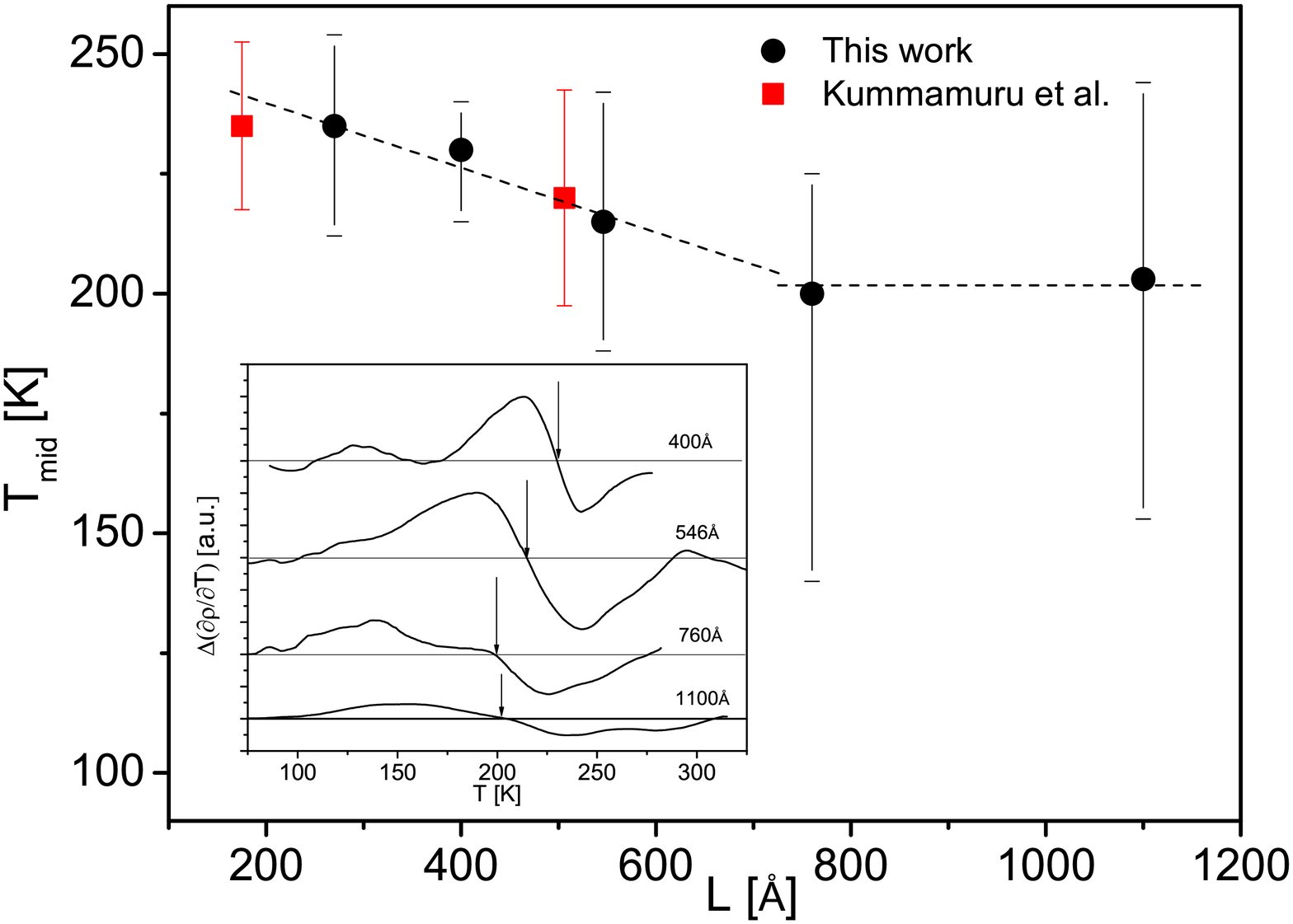}
\caption{(Color online) Transition temperature as a function of film thickness. Red squares correspond to data in ref..\cite{kummamuru} The vertical bars are not error bars but the width of the hysteretic region (see text). The dashed lines are guides to the eye. Inset: Difference between the derivatives of $\rho (T)$ curves for cooling and warming cycles of films with increasing thicknesses. For clarity, the vertical zero of each curve (horizontal lines) has been arbitrarily shifted. The arrows indicate the temperature $T_{mid}$ (see text).}
\label{tmid}
\end{figure}

To appreciate the mentioned behavior, we plot in 
Fig.\ref{tmid} $T_{mid}$ as a function of film 
thickness $L$. Note that the bars are not error bars but 
represent the irreversibility region width, $\Delta T$, as defined 
above. It can be observed that there exists a systematic shift of the 
hysteretic region towards lower temperatures as the film thickness is increased. In fact, $T_{mid}$ goes down and tends to saturate as the film thickness becomes larger than 750\,\AA. For comparison, we include the data from Kummamuru et al. 
\cite{kummamuru}(squares). As can be seen, their data are in very good 
agreement with our results. This would be expected because both set of 
films have been grown following the same procedure and over the same type of substrate. In the data of Fullerton et al.\cite{fullerton} a similar decreasing trend for $T_{mid}$ may be observed,
but a much steeper decrease of $T_{mid}$ with thickness seems to be 
present, at least for the samples reported (63\AA \- and 200\AA). Since it is well known that the magnetism at Cr surfaces strongly depends on conditions at its surface or interface \cite{zabel}, this different behavior might be related to the role played by the CrMn layers in their Cr/CrMn superlattices.\cite{fishman}

Other important features are worth to remark. In the inset of Fig.
\ref{tmid} we plot the difference of the derivatives as a 
function of $T$ for films of increasing thickness. For the sake of 
clarity, the vertical zero of each curve has been shifted by an 
arbitrary offset. The important point here is not the absolute value of $\partial\rho/\partial T$, but the temperature $T_{mid}$ at which $\partial\rho_c/\partial T$ (cooling) equals $\partial\rho_w/\partial T$ (warming).
Note that $\Delta T$ increases,  
while the amplitude of the difference between $\partial\rho_c/\partial T$ and 
$\partial\rho_w/\partial T$, measured at $T_L$ and $T_H$, decreases. These two last observations give a clear clue on why the irreversible behavior in $\rho(T)$ is not likely to be seen in thicker samples.

If we identify $T_{mid}$ as the temperature at which a first order 
transition takes place between a state with an ISDW with $N$ nodes to a 
new state with $N$+1 nodes when the temperature is reduced, then our data clearly show that the transition temperature {\it decreases} as the film thickness {\it is increased}, tending to saturation for films thicker than 750\,\AA. In order to explain this behavior of $T_{mid}(L)$, let us look more in detail the quantization of the ISDW in thin films, which is important to understand the hysteresis of the resistivity in cooling-warming cycles.

As a product of different internal interactions, $\delta$ depends on
temperature.\cite{werner} It varies quite linearly from, 
$\delta$\,=\,0.0379  at $T$\,=\,300\,K, to $\delta$\,=\,0.044 at $T$\,=\,200\,K. As a direct consequence, $\Lambda_{ISDW}$ changes from 76\,\AA \- to 65.5\,\AA \- in the
same temperature range. At much lower temperatures it saturates to a 
constant value $\delta_c$\,=\,0.048 giving $\Lambda_{ISDW}$\,$\approx$\,60\,\AA .

Since the magnetic instability in the electron system depends on the combination
$\cos(\mathbf{Q}_{+}\cdot\mathbf{r})+\cos(\mathbf{Q}_{-}\cdot\mathbf{r})$,
it is clear that the resulting beating has a fast variation modulated
by a long wavelength $(Q_+ - Q_-)/2 =
\frac{2\pi}{a/\delta}$. Thus, the ISDW in Cr has a wavelength
\begin{equation}
\Lambda_{ISDW}=\frac{2.88}{\delta}\text{\AA}.
\label{lambda}
\end{equation}

\begin{figure}[t]
\includegraphics[width=0.45\textwidth]{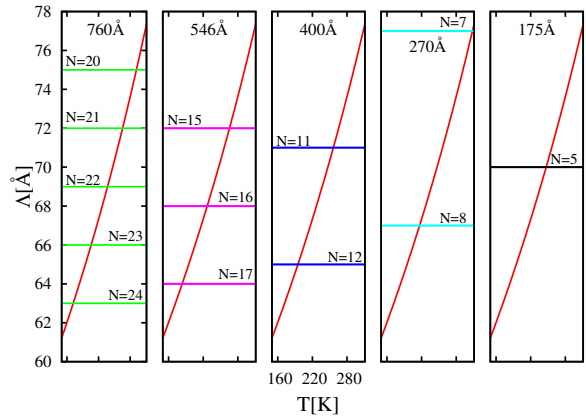}
\caption{(Color online) Available ISDW modes for different film thicknesses in the temperature region of interest. The temperature scale is the same for all panels.}
\label{mod3}
\end{figure}

When the ISDW is confined in Cr films with non-magnetic interfaces, as is the case in our Cr films, we expect a
configuration with antinodes (or nodes)\cite{mibu} at the films' 
surfaces. For this case,
the available modes are separated by half a wavelength. Given a film with thickness $L$, the allowed values
for $\Lambda_N$ are therefore $2 L / N$, where $N$ is either the number of nodes inside the film (antinodes at the surfaces), or antinodes in the film (nodes at the surface). 

In the window of wavelength values found in bulk chromium given by $\delta(T)$ (60\,\AA \- to 78\,\AA), there are only a few modes available for each film thickness. Their number increases as $L$ increases. This is shown in Fig.\ref{mod3} where we plot the allowed modes for each film thickness. The lines correspond to $\Lambda (T)$ as calculated from equation \ref{lambda}, using a linear approximation for $\delta (T)$ between 200\,K and 300\,K, obtained from the data in Ref.\cite{werner}. This approximation seems reasonable because in the thickness range in which our samples lay (270-1100\,\AA ) we do not see appreciable 
variations of the N\'eel temperature, indicating that there is no strong 
dependence of $\delta$ with thickness.\cite{fishman-rev,rotenberg}
 
As can be seen in Fig.\ref{mod3}, in the the film\cite{kummamuru} 
of 175\,\AA, the transition occurs between N\,=\,4 and N\,=\,5 at the 
temperature where the line corresponding to $\Lambda_5 \approx 
\Lambda_{ISDW}(T)$. Correspondingly, for the 270\,\AA \- film the 
transition between N\,=\,7 and N\,=\,8 lies at a lower temperature. 
From this, it seems that the transition temperature would decrease with 
$L$. This reasoning, however, cannot be extended to the other film 
thicknesses based on the information displayed in Fig.\ref{mod3} alone, because more modes than one become available for the ISDW, and there is no way to choose {\it a priori} at which one the transition would occur. In order to understand how the transition temperature depends on $L$, in Section \ref{model} we propose a phenomenological model which takes into account this variation of the number of permitted modes when $L$ increases, and we also incorporate some dynamics.

\section{Model}
\label{model}
\subsection{The Free Energy of the ISDW}

When the temperature is decreased below $T_N$, the wavelength of the ISDW in the Cr film would try to follow it's bulk
behavior.\cite{werner} However, as discussed in the previous section: the ISDW is confined in a film of a given thickness, the
variation in temperature of $\Lambda_{ISDW}$ is quantized, and the number of nodes varies discontinuously
from $N$ to $N$\,+\,1, in a process that involves a first order phase transition. Neutron diffraction\cite{fullerton} as well as X-ray experiments\cite{kummamuru} confirm this physical picture in Cr/CrMn multilayers and in Cr thin films respectively.

Let us consider a one-dimensional model of a transverse ISDW with ${\mathbf Q}$\,=\,$q \hat{z}$, where $\hat{z}$ is the direction perpendicular to the film's surface. In this case the order parameter takes the form $\Psi$\,=\,$\Psi_0 e^{iq z}$. We can then
write an expression for the free energy (per unit length) depending on $\Psi_0$ and $q$,

\begin{eqnarray}
F[\Psi_0,q ] &=&F_0 + a(T-T^{bulk}_N)|\Psi|^2 + b|\Psi|^4 \nonumber\\
&+& c|(\nabla - iq_0)\Psi |^2 + U_{surface},  \label{free_en}
\end{eqnarray}

where $F_0$ is the part of the free energy which is independent of the ISDW, $T^{bulk}_N$\,=\,311\,K and $q_0$ is the wave vector that the ISDW
would have in the bulk material. Following McMillan\cite{mcmillan} we include the term $|(\nabla - iq_0)\Psi |^2$ which gives a
contribution proportional to $(q-q_0)^2\Psi_0^2$. This takes into
account the fact that there exists an energy cost in order to deform the ISDW
with respect to it's bulk form. The key point is that $q_0$ varies in temperature
and these variations should be followed by the ISDW in order to keep the
energy at a minimum.

There is, nevertheless, a competition between the energy paid to
deform the ISDW respect to the bulk, and the energy cost for not
satisfying the boundary conditions imposed by the surface. As already mentioned, the most natural way to choose these boundary conditions for Cr films that are not in contact with magnetic materials is to have nodes or antinodes at the surfaces. For both cases $\Lambda_N =2 L / N$, i.e. $q =\pi N / L=k_N$. Considering this, the surface contribution to the free energy is given by a term
proportional to $(q-k_N)^2\Psi_0^2$. The way to account for the
different possible available modes $\{k_N\}$ is simply by selecting the one that gives the lowest contribution to the free energy
at a given temperature (see Fig.\ref{mod3}). Then, we explicitly have for the free energy,

\begin{eqnarray}
F[\Psi_0,q ] &=&F_0 + a(T-T^{bulk}_N)\Psi_0^2 + b \Psi_0 ^4 \nonumber\\
&+& c(q - q_0)^2 \Psi_0 ^2 + U\min_{\{k_N\}}[(q-k_N)^2\Psi_0^2].
\label{free_en2}
\end{eqnarray}

\begin{figure}[tbp]
\includegraphics[width=8.5cm]{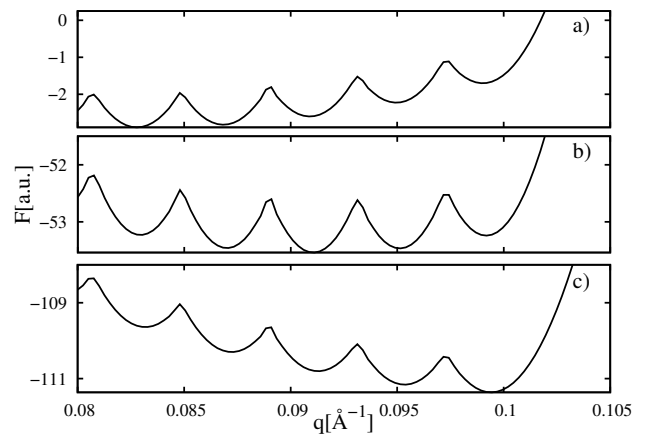}
\caption{Cuts of the two dimensional free energy, $F[\Psi_0,q(T)]$, at different temperatures and for a fixed value
of $\Psi_0$. a) $T/T_{N}=0.96$, b) $T/T_{N}=0.74$, c) $T/T_{N}=0.48$. Available modes correspond to a film of $760$\,\AA.}
\label{mod4}
\end{figure}

If the nodes (or antinodes) are pinned at the surface (enough large $U$), then the free energy
displays a series of minima corresponding to parabolic potentials
sitting close to each $k_N$. This is shown in Fig.\ref{mod4} for a fixed
amplitude of the order parameter $\Psi_0$ and adequate numeric
parameters. Note the change in the potential landscape with temperature. As $T$ is decreased the barriers between modes increase and the free energy changes in such a way to give preference to the modes with higher $q$'s, i.e. shorter wavelengths and more nodes (antinodes) in the ISDW.

In this simple way we can understand how the ISDW's
wavelength is jumping from $N$ to $N$\,+\,1 nodes and so on as the temperature
decreases. Given the fact that each of these jumps involves a first order
transition, hysteresis is expected in cooling-warming cycles.

\subsection{Monte-Carlo Simulation Results}

\begin{figure}[tbp]
\includegraphics[width=8.5cm]{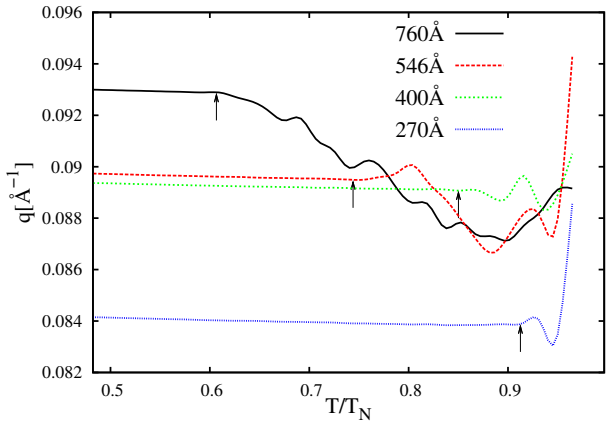}
\caption{(Color online) Average $q$ as a function of temperature for different film thicknesses. The fluctuations in $q$ at high temperatures freeze to a fixed number of nodes at a temperature $T_{mid}$ which depends on the film thickness (arrows).}
\label{mod5}
\end{figure}

\begin{figure}[t!]
\includegraphics[width=8.5cm]{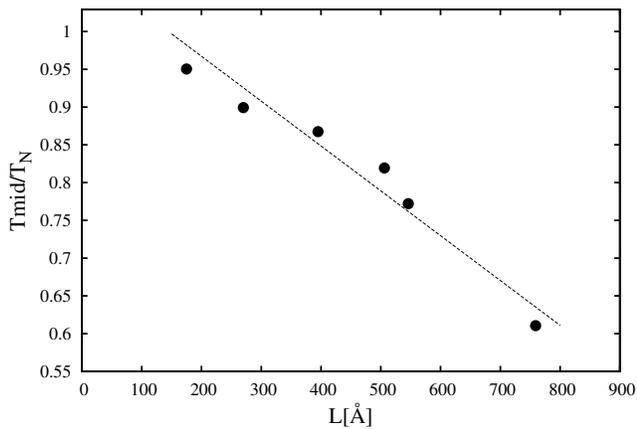}
\caption{Freezing temperature from Fig.\ref{mod5} as a function of film thickness. The dashed line is a guide to the eye.}
\label{mod6}
\end{figure}

One of the most important results of this work is the description of 
how the temperature of the maximum hysteresis, $T_{mid}$, moves to 
lower values as the thickness of the Cr films is increased. From the 
scheme shown in Fig.\ref{mod3} it can be inferred that when many ISDW modes are available (larger $L$'s), a large barrier should be needed in order to freeze the state of the system with the largest number of nodes, otherwise the system would wander between all available modes. Since the ISDW's amplitude $\Psi_0$ increases with decreasing temperature as well as the barriers do, lower temperatures are needed to freeze the system in a high $k_N$ mode.

To better quantify this, we introduce some dynamics by means of a 
Monte-Carlo simulation of the system. Here we have a two dimensional 
phase space $(q,\Psi_0)$ and an energy given by Eq.\ref{free_en2}. 
The resulting average value of $q$ after several runs is shown in Fig.\ref{mod5} for different film thickness as the temperature is decreased. At high temperatures, the system explores all possible $k_N$'s for each $L$, but as the temperature is lowered, the ISDW chooses only one $k_N$ and stays there no matter how low the temperature goes. This behavior defines a freezing temperature that we identify with $T_{mid}$.

The values obtained from the results in Fig.\ref{mod5} for the 
freezing temperature corresponding to each film thickness, are plotted 
in Fig.\ref{mod6}. Clearly, the freezing temperature follows the 
same trend as the experimental data for $T_{mid} (L)$, i.e., it goes 
down as the film thickness is increased. The saturation seen in the 
transition temperature in our data for $L$'s larger than 750\,\AA \- 
naturally occurs in our Monte-Carlo simulations if we put, mimicking 
the experimental results,\cite{werner} a constant $\delta (T)$ at low 
temperatures. With this simple phenomenological model we do not pretend 
to get a quantitative agreement with our experimental data, but it is 
important to understand the physics behind the observed hysteretic phenomena.

\section{Conclusions}

We have shown that the hysteretic behavior that appears in $\rho (T)$ in epitaxial Cr thin films is a consequence of the quantization of a transverse incommensurate spin density wave in the direction perpendicular to the film surfaces, with a first order phase transition occurring between domains with $N$ and $N$\,+\,1 nodes (or antinodes). The number $N$ at which the transition manifests is film thickness dependent, being larger for thicker films. An analysis of the one dimensional ISDW free energy, estimated using McMillan's approach for CDW's, together with some dynamics introduced via Monte-Carlo simulations, show that the potential barriers between allowed modes increases as the system is cooled down. As a consequence of this, and for each $L$, at high temperatures the system explores all available modes but remains frozen in a high $k_N$ state at low enough temperatures. Therefore, this simple phenomenological model explains in a natural way the experimental results which show that the temperature at which the transition between $N$ and $N$\,+\,1 modes takes place, decreases as the film thickness is increased.

\section{Acknowledgements}

We would like to thank H. Pastoriza and K. Ogando for help with the patterning of the samples. E.O., E.K. and C.A.B. are members of CONICET. L.T. has a scholarship from CONICET. Work partially supported by PIP No.11220080101821 CONICET, PICT R1776 ANPCyT, and PIP No.1122008010111001 CONICET.

\end{document}